\documentclass[fleqn,usenatbib]{mnras}

\usepackage{newtxtext,newtxmath}
\usepackage[T1]{fontenc}
\usepackage{color,xcolor}
\usepackage{soul}
\setstcolor{red}
\usepackage{float}
\usepackage{caption}
\usepackage{subcaption}

\DeclareRobustCommand{\VAN}[3]{#2}
\let\VANthebibliography\thebibliography
\def\thebibliography{\DeclareRobustCommand{\VAN}[3]{##3}\VANthebibliography}

\usepackage{graphicx}	
\usepackage{amsmath}	

\usepackage{xcolor}
\usepackage{hyperref}

\title[Radio sensitivity of MSPs in the Sgr dSph]{Radio sensitivity to a new population of millisecond pulsars in the Sagittarius Dwarf Spheroidal Galaxy}

\author[L. Gebauer W.  et al.]{
Lucia Gebauer Werner,$^{1,2}$\thanks{E-mail: lgebauerw@gmail.com}
Oscar Macias,$^{3,1}$
Christoph Weniger$^{1}$
\\
$^{1}$GRAPPA Institute, University of Amsterdam, Science Park 904, 1098 XH Amsterdam, Netherlands\\
$^{2}$ Max-Planck-Institut für Radioastronomie, Auf dem Hügel 69, D-53121 Bonn, Germany\\
$^{3}$ Department of Physics and Astronomy, San Francisco State University, San Francisco, California 94132, USA\\
}

\date{Accepted XXX. Received YYY; in original form ZZZ}

\pubyear{2024}

\begin{document}
\label{firstpage}
\pagerange{\pageref{firstpage}--\pageref{lastpage}}
\maketitle

\begin{abstract}

Observations with the Fermi Gamma-Ray Space Telescope reveal an excess of extended gamma-ray emission likely caused by an undiscovered population of millisecond pulsars (MSPs) in the core of the Sagittarius dwarf spheroidal galaxy (Sgr dSph). However, additional evidence, such as multi-wavelength searches, is necessary to confirm this theory. A significant discovery could be made if radio pulsations from individual MSPs in the Sgr dSph are detected. In this study, we investigate the possibility of detecting MSPs in the Sgr dSph with present and upcoming radio surveys using a phenomenological model based on the observed luminosity function of MSPs in the Milky Way's globular clusters. Our findings suggest that the Square Kilometer Array (SKA) is the most sensitive instrument for detecting these objects. 
We demonstrate that to observe one MSP with MeerKAT, we would need to perform a pointing observation of the core region of the Sgr dSph for about two hours. In this same observation time, SKA can identify $9^{+5}_{-3}$ MSPs in the entire system. Based on the distance of the Sgr dSph galaxy and our dispersion measure distance estimate, we find it possible to differentiate between MSPs belonging to the Sgr dSph and those of the Galactic disk and bulge. Furthermore, the MSPs hypothesis for the Sgr dSph gamma-ray excess could be confirmed at the 99.7\% confidence level by detecting at least six MSPs in a two-hour SKA observation of the Sgr dSph. 

\end{abstract}

\begin{keywords}
Sgr dSph  -- Millisecond Pulsars -- MeerKAT -- SKA 
\end{keywords}



\section{Introduction}\label{sec:intro}
More than a decade ago, \citet{Su2012} claimed the detection of two counter-propagating features from the center of the galaxy within the Fermi bubbles (which are colossal lobes of gamma radiation emanating from the Galactic center) using Fermi-LAT gamma-ray observations~\citep{Ackermann_2014}. Due to their spatial characteristics, these jet-like structures were initially believed to be connected to past energetic activity from the supermassive black hole Sgr A$^\star$~\citep{Su2012}.
While the jet-like substructure in the northern bubble has not been confirmed, a subsequent study by the Fermi-LAT collaboration did confirm the detection of a bright large-scale elongated structure in the southern part of the Fermi bubbles called the ``Fermi cocoon''~\citep{Ackermann_2014}.

The quest to uncover the origins of the Fermi cocoon stands as one of the most important challenges in the field of high-energy astrophysics. Recently, \citet{Crocker2022} demonstrated a significant spatial correlation between the Fermi cocoon and the stellar population of the Sgr dSph galaxy. They did so by comparing Fermi-LAT data with the distribution of RR Lyrae stars from Gaia Data Release 2 \citep{Vasiliev_2020a}.

\citet{Crocker2022} also studied how systematic uncertainties in the Galactic diffuse gamma-ray emission impact the results. They found that regardless of the alternative diffuse emission models considered, the stellar template was detected with a significance of over $\geq 8.1\sigma$. Moreover, the probability of the Fermi cocoon and Sgr dSph spatially overlapping just by chance was less than 1\%.
Detecting the Sgr dSph in gamma-ray data is not entirely surprising considering its significant stellar mass and proximity to the Sun. The Sgr dSph is the celestial body with the highest stellar mass per distance square ($M_\odot/d^2$) that had not been detected in gamma rays. 

Interestingly, the H.E.S.S collaboration searched for extended gamma-ray emission from the Sgr dSph at TeV-scale energies but only found a minor fluctuation with a significance of $2.05\ \sigma$ \citep{Abramowski_et_al_2014}. A sensitivity analysis conducted by \citet{Viana_et_al_2012} in the same system looked for potential sources of $\gamma$-ray emission detectable with Cherenkov telescopes. They concluded that the predicted $\gamma$-ray emission from millisecond pulsars would outshine the expected $\gamma$-ray signal from dark matter annihilation by several orders of magnitude. Additionally, \citet{Evans:2022zno} recently detected various gamma-ray point source candidates overlapping with the positions of globular clusters of the Sgr dSph.

Given their robust spatial correlation, how can the stellar map of the Sgr dSph be causally connected to the Fermi cocoon gamma-ray map?

The Sgr dSph galaxy is one of the closest and most massive satellites of the Milky Way, with a current stellar mass of around $10^8 M_\odot$ \citep{Vasiliev_2020a, Vasiliev_2020b}. Due to the Galaxy's gravitational pull, the Sgr dSph is currently being disrupted, and elongated streams of stars are being tidally stripped from the dwarf as it is absorbed into the Milky Way~\citep{Belokurov_2006}. Despite this gravitational disruption, the Sgr dSph galaxy still has a core and an elongated stellar stream that extends over 100 kpc~\citep{Majewski_2003, Law_2004}. This core is located at (RA, DEC) = $(283.76^\circ, -30.48^\circ)$ in equatorial coordinates and is approximately 26.5 kpc away from Earth.  

One potential mechanism for extended gamma-ray emission from the Sagittarius dwarf spheroidal galaxy (Sgr dSph) is hadronic radiation\citep{Strong_2010} resulting from the collision between energetic cosmic rays and interstellar gas. However, due to the tidal ram pressure of the Milky Way, the Sgr dSph has lost most of its gas material, which makes the possibility of hadronic gamma-ray radiation unlikely. Moreover, star formation in the Sgr dSph ceased about $\sim 2$ to 3 billion years ago~\citep{Siegel_2007}, further weakening the motivation for hadronic gamma-ray radiation given the absence of recent core-collapse supernovae. 
Based on our current knowledge of the Sgr dSph, the potential emission sources can be narrowed down to two possibilities. The first is the emission from an unresolved population of millisecond pulsars in the Sgr dSph region. The second possibility is dark matter self-annihilations.

Recently, \citet{Venville2023} conducted a detailed study on the infall of Sgr dSph into the Milky Way by utilizing the results of advanced hydrodynamical simulations of the system \citep{Tepper-Garcia_2018}. They considered the dark matter, stellar, and gaseous components of both the satellite and the host galaxy. The simulation was run for a total of 3.6 Gyr, during which the dark matter halo of Sgr dSph underwent significant tidal disruption. This allowed \citet{Venville2023} to obtain an accurate estimate of the J-factor (i.e., the line-of-sight integration of the square of the dark matter density, which scales linearly with the dark matter emissivity) and thus evaluate the dark matter emission hypothesis for the Sgr dSph.

They concluded that the dark matter annihilation cross-section required to explain the observed $\gamma-$ray emission from Sgr dSph was incompatible with existing constraints. As a result, the most likely explanation for the Sgr dSph excess is the millisecond pulsars (MSPs) hypothesis.

\citet{Crocker2022} demonstrated that a population of MSPs could account for the signal's spatial morphology, luminosity, and spectrum. The spectrum was found to be a combination of prompt magnetospheric radiation and inverse Compton scattering of cosmic microwave background photons by $e^{\pm}$ pairs injected by the MSPs.

MSPs emit across the electromagnetic spectrum, which enables us to search for evidence of undiscovered MSPs in the Sgr dSph through multi-wavelength data. In this article, we investigate the sensitivity of current and future radio telescopes (such as MeerKAT and SKA) to detect MSPs in the Sgr dSph. For this, we use the measured radio luminosity function of MSPs in globular clusters and include the latest instrumental response of these telescopes.

Specifically, Section \ref{sec:method_GC_to_Sgr} describes how we scaled data from globular clusters to estimate the number of MSPs in Sgr dSph and the potential number of observable radio MSPs. In Section \ref{sec:RadioTelescopes} we provide details of the telescope parameters we will use as an input in the sensitivity equation to estimate the telescope capacities. In Section \ref{sec:Results}, we present our findings on whether it is possible to detect MSPs in the Sgr dSph by using MeerKAT and SKA. We also provide details of our estimates on the possibility that some of these MSPs are foreground MSPs in the line-of-sight of the Sgr dSph. Finally, we have summarized our research findings and drawn our conclusions in Section \ref{sec:DiscussionConclussions}.

\section{The Luminosity Function of MSPs in the Sagittarius dwarf spheroidal}\label{sec:method_GC_to_Sgr}
To estimate the number of MSPs in Sgr dSph, we use a procedure that was previously applied to the galactic center, as described in \citet{Calore_2016}. The procedure involves building a model based on observations of radio emission from MSPs in globular clusters of the Milky Way. Specifically, we estimate the surface density of radio-bright MSPs, defined as MSPs with a flux density of 0.01 mJy at a frequency of 1.4 GHz \citep{Bagchi_2011}. 

In our work, we estimate the luminosity of MSPs in the Sgr dSph by analyzing data obtained from radio millisecond pulsars in globular clusters and their gamma-ray emission. This involves rescaling the data to a distance of 26.5 kpc \citep{Vasiliev_2020a}. However, we should exercise caution using this method as it assumes that the MSPs in globular clusters and those in Sgr dSph have similar emission properties. If their evolutionary histories differ, then this method may not be accurate. Nonetheless, this approach can give us a rough estimate of the dwarf's emission.

As suggested by \citet{Calore_2016}, we utilize a data-driven luminosity ratio to describe the populations of MSPs in globular clusters:
\begin{equation}
    \label{eq:GammaRadioRatio}
    \mathcal{R_\text{rb}^\gamma} \equiv 
    \frac{\left< L^\text{Sgr dSph}_\gamma\right>}{\left<N^\text{Sgr dSph}_\text{rb}\right>} \simeq \frac{L^\text{stacked}_\gamma}{N^\text{stacked}_\text{rb}}
\end{equation}
where $L^{\rm stacked}_{\gamma}$ is the net gamma-ray luminosity of the globular cluster, and $N^{\rm stacked}_{\rm rb}$ is the total number of radio bright MSPs (i.e., with flux densities greater 0.01 mJy at 1.4 GHz) in a given globular cluster. It's worth noting that the $\gamma$-ray and radio emission for MSPs do not necessarily coincide---therefore, there could be $\gamma$-ray only MSPs and vice versa. This means we cannot reliably estimate the average gamma-ray luminosity of radio-bright MSPs using $\mathcal{R}_\mathrm{rb}^{\gamma}$. However, it can reasonably estimate a potentially large population of MSPs in the Sgr dSph. It is also important to mention that we are studying the average emission properties, so the gamma-ray luminosity function details are irrelevant in this case.

\subsection{Stacked Gamma-ray Luminosity}\label{subsec:Stacked-Gamma-ray-Luminosity}
Millisecond pulsars are believed to be the leading cause of gamma-ray emission from globular clusters. The spectra of these clusters contain two distinct emission processes: curvature radiation (CR) from the MSP magnetosphere and inverse Compton (IC) emission from relativistic pairs launched into the environment by MSPs. A recent study by \citet{Song_2021} analyzed Fermi-LAT gamma-ray observations from different globular clusters of the MW. That study was able to disentangle the CR component from the IC component in a number of the observed globular clusters.

We use the results of \citet{Song_2021} to estimate the fraction of an MSP's luminosity that goes into the gamma-ray CR component compared to the luminosity released in the radio band.  We rely on the findings presented in \citet{Song_2021}, which used an exponential cutoff fit to estimate the power released in curvature radiation. The relevant data for the globular clusters analyzed by them is summarized in Table~\ref{tab:GC_info}.

\begin{table*}
    \centering
    \begin{tabular}{|l c c c c c |}
        \hline
        Globular Cluster & $d$ [kpc] & $L_{\gamma}\ [10^{34}\text{erg s}^{-1}]$ & $N_\text{rad}$ & $N_\text{obs}$ & $N_\text{rb}$\\
        \hline 
        Ter 5 & 6.9 & $27.46 \pm 1.83$ & $82\pm 16$ & 25 & $2.2 \pm 0.4$ \\
        47 Tuc & 4.5 & $4.67 \pm 0.30$ & $37\pm 10$ & 14 & $0.9 \pm 0.3$\\
        M15 & 5.5 & $2.38 \pm 0.62$ & $41\pm 15$ & 9 & $1.0 \pm 0.4$\\
        NGC 6440 & 8.5 & $5.12 \pm 0.97$ & $48\pm 21$ & 6 & $1.2 \pm 0.5$ \\
        NGC 6752 & 4.0 & $0.33\pm 0.08$ & $21\pm 10$  & 5 & $0.6 \pm 0.3$ \\
        M5 & 7.5 & $0.52 \pm 0.24$ & $13\pm 6$ & 5 & $0.3 \pm 0.1$ \\
        \hline
        Stacked & & $40.5\pm 21.9$ & $242\pm 34$ & 64 & $5.0 \pm 0.7$\\
        \hline
    \end{tabular}
    \caption[]{
    Gamma-ray luminosity and number of radio detected MSPs in 6 globular clusters. The values for the MSPs can be found in the list of Pulsars in Globular Clusters \footnote{\url{https://www.mpifr-bonn.mpg.de/staff/pfreire/GCpsr.html}}. The number of radio MSPs, $ N_\mathrm{rad} $, and the number of observed MSPs $ N_\mathrm{obs} $ are from \citet{Bagchi_2011}. We show the $ N_\mathrm{rad} $ values obtained with the luminosity fit of model 3 using $ N_\mathrm{obs} $ listed in this table. Lastly, we show the stacked gamma-ray     luminosity $ L^\mathrm{stacked}_\gamma$, the stacked number of radio MSPs $ N^\mathrm{stacked}_\mathrm{rad}$, and the stacked number of radio-bright MSPs, $ N^\mathrm{stacked}_\mathrm{rb} $, using Model 3 ($ \mathcal{N}(\mu =-1.1, \sigma=0.9) $). The uncertainties for the stacked values are the sum of the squared estimate of errors.}
    \label{tab:GC_info}
\end{table*}

In order to calculate the stacked gamma-ray luminosity $L^\mathrm{stacked}_\gamma$, we estimate the CR emission from globular clusters Terzan 5, 47 Tucanae (NGC 104), M15 (NGC 7078), NGC 6440, NGC 6752, and M5 (NGC 5904) following \citet{Song_2021}.
For some globular clusters, the analysis conducted by \citet{Song_2021} did not have enough statistics to distinguish the CR component from the IC component. This was particularly true for M15 (NGC 7078). In such situations, the system's estimated total gamma-ray luminosity is utilized.

\begin{table*}
    \centering
    \begin{tabular}{lccccc}
       \hline
       Luminosity function $(\mu, \sigma)$ & $N^\text{stacked}_\text{rad}$ & $N^\text{stacked}_\text{rb}(d\simeq 26.5\text{ kpc}$) & $\mathcal{R}_\text{rb}^\gamma$[$10^{34}$erg s$^{-1}$]& $N_\text{rb}^\text{Sgr}$ & $N_\text{rad}^\text{Sgr}$\\
       \hline
       Model 1 ($\mu=-1.1$, $\sigma = 0.9$) & $473 \pm 70$& $6.2\pm 0.9$&  $6.49 \pm 3.6$&  $25 \pm 15$& $1928 \pm 1167$\\
       Model 2 ($\mu=-0.61$, $\sigma = 0.65$) & $307 \pm 45$& $3.6 \pm 0.51$&  $11.1 \pm 6$ &  $14\pm 9$& $1251 \pm 758$\\
       Model 3 ($\mu=-0.52$, $\sigma = 0.68$) & $242 \pm 34$& $5.0\pm 0.7$& $8.12\pm 4.5$&  $20 \pm 12$& $986\pm 555$\\
       \hline
    \end{tabular}
    \caption[]{Number of radio MSPs per model. The first column shows the number of stacked radio MSPs from \citet{Bagchi_2011}, $N^\mathrm{stacked}_\mathrm{rb}$ is the number of stacked MSPs rescaled to Sgr dSph distance, $\mathcal{R}^\gamma_\mathrm{rb}$ is the gamma-ray radio-bright ratio in $10^{34}\mathrm{erg\ s}^{-1}$, $N^\mathrm{Sgr}_\mathrm{rb}$ is the number of radio-bright MSPs from Sgr dSph and $N^\mathrm{Sgr}_\mathrm{rad}$ is the total radio MSPs in the dwarf.}
    \label{tab:results_per_model}
\end{table*}

\subsection{Radio Luminosity Function}
\citet{Bagchi_2011} conducted Monte Carlo simulations to model the luminosity distribution of recycled pulsars in globular clusters. That study found that a log-normal model generally better matched the data. Following their approach, we have selected a subset of their globular cluster sample based on the availability of information on the flux at 1.4 GHz and the measured gamma-ray luminosities. 

To estimate the number of radio-bright MSPs, we model the MSPs' luminosity function with a log-normal distribution written as:

\begin{equation}
    f_\text{log-normal}(L_\nu) = \frac{\log_{10}e}{L_\nu}
                \frac{1}{\sqrt{2\pi \sigma^2}}
                \exp \left[ \frac{-(\log_{10}L_\nu - \mu)^2}{2\sigma^2}\right]
\end{equation}
where the $\mu$ and $\sigma$ are the mean and standard deviation of the distribution respectively, and the ``pseudo-luminosity'' $L_\nu$ can be approximated as  $L_\nu = S_\nu d^2$ since it is typically difficult to determine some of the luminosity parameters.

There are two main numerical methods for determining the parameters of the pulsar luminosity function \citep{Bagchi_2011}. The first is the full dynamical approach, where a simulation is performed by seeding a model galaxy with pulsars according to various birth locations and rotational parameters. These synthetic pulsars are evolved kinematically and rotationally and then compared to the observed sample. The second approach is the snapshot method, where pulsars are seeded at their final positions in the galaxy, forming a picture of the present-day population without assumptions about their spin-down or kinematic evolution.

In their study, \citet{Bagchi_2011} investigated the log-normal function using three different models. Model 1 was based on the dynamical pulsar model proposed by Faucher (2006), which was used to model galactic isolated radio pulsars. A luminosity fit was obtained with $\mu=-1.1$ and $\sigma = 0.9$. On the other hand, Model 2 and Model 3 were both data-driven models. Model 2 was selected based on the Kolmogorov-Smirnov (KS) statistic to fit the observed globular cluster data, resulting in a maximum KS statistic of $\mu=-0.61$ and $\sigma = 0.65$. Meanwhile, Model 3 was selected based on the minimum $\chi^2$, which had $\mu = -0.52$ and $\sigma=0.68$. Table~\ref{tab:results_per_model} shows the number of radio sources estimated using the radio luminosity function for each model.

In this paper, our focus will be on the results obtained from Model 3. However, it is worth mentioning that Table~\ref{tab:results_per_model} shows that the other two models provide optimistic and pessimistic scenarios for the number of radio-bright MSPs in the galaxy. We will discuss these results later. As pointed out in \cite{Calore_2016}, the log-normal radio luminosity function has a high-luminosity tail that predicts sources brighter than the brightest MSPs detected in globular clusters. To avoid a bias towards excessively bright sources, we have truncated the radio luminosity function to a maximum pseudo-luminosity of 30 mJy kpc$^2$. When we rescale it to the distance of Sgr dSph, this translates to a flux density of 0.04 mJy.

\subsection{Rescaling the luminosity function of MSPs in globular clusters to the Sagittarius dwarf galaxy}

We estimate that the ratio between the stacked gamma-ray luminosity of our globular cluster sample ($L^\text{stacked}_\gamma$) and the number of radio-bright MSPs is given by
\begin{equation}
    \mathcal{R}_\text{rb}^\gamma \simeq (8.12\pm 4.5)\times 10^{34} \;\text{erg s}^{-1},
\end{equation}
where the total number of radio-bright MSPs was calculated using the radio-luminosity function of MSPs in globular clusters, assuming Model 3 in \citet{Bagchi_2011}.
We can now use Eq.~\ref{eq:GammaRadioRatio} to calculate the number of radio-bright MSPs in the Sgr dSph, based on the prompt gamma-ray luminosity measured in \citet{Crocker2022}.

It is worth noting that the gamma-ray luminosity in \citet{Song_2021} was calculated for energies greater than 300 MeV, while in \citet{Crocker2022} it was assumed to be valid from 500 MeV. Therefore, we use the best fit spectrum reported in \citet{Crocker2022} and assume that it holds true until 300 MeV, which is a reasonable approximation. We obtain that the prompt gamma-ray luminosity emitted by the Sgr dSph for energies greater than 300 MeV is
\begin{equation}
    L_\text{Sgr dSph} = (1.65\pm 0.38) \times 10^{36} \text{erg s}^{-1}
\end{equation}
Taking the ratio between $L_\text{Sgr}$ and $\mathcal{R}_\text{rb}^\gamma$, we then obtain that the expected number of radio bright MSPs in the Sgr dSph, as predicted by Model 3 in \citet{Bagchi_2011}, is

\begin{equation}
    N^\text{Sgr}_\text{rb} \simeq 20 \pm 12.
\end{equation}

The results of the estimates are summarized in Table~\ref{tab:results_per_model}. We have calculated the number of radio-bright MSPs in the stacked globular clusters, assuming they are at the same distance as the Sgr dSph. We used our reference luminosity function normalized to the number of radio pulsars, as indicated in Table~\ref{tab:GC_info}. Additionally, we show the results for the two other luminosity functions from \citet{Bagchi_2011}, which bracket the uncertainties implied by the observed MSPs. Although the total number of radio MSPs is uncertain by at least a factor of a few, the number of radio-bright MSPs is much better constrained, since direct observations support it.

As seen in Table 2, the count of radio-bright sources slightly depends on the utilized radio luminosity function. However, to simulate sources in the Sgr dSph, we require the total number of all radio MSPs. Thus, for the remaining part of this article, we will adopt ``Model 3,'' which gives us a total count of radio MSPs of $N_\text{rad}^\text{Sgr} = 986.4 \pm 555.7$.

\subsection{Sgr dSph Surface Density}\label{subsec:SurfDensity}

In order to create a map that shows the surface density of radio-bright MSPs in the Sgr dSph, we need to obtain a model that describes the MSPs' spatial distribution in the dwarf, the value of $\mathcal{R}_\mathrm{rb}^\gamma$, and the energy intensity. To get the spatial distribution model of MSPs in the Sgr dSph, we can use the spatial distribution of RR Lyrae stars as a proxy since they also belong to the old stellar populations and are a suitable proxy for this purpose. In contrast, in the context of the Galactic center gamma-ray excess~\citep[e.g.,][]{Goodenough:2009gk, Hooper:2011ti, Calore:2014xka, Fermi-LAT:2015sau, Fermi-LAT:2017opo, Macias:2016nev}{}{}, analytical models of the distribution of dark matter were used, as described in \citet{Calore_2016}.
\begin{figure}
    \centering
    \includegraphics[scale=0.35]{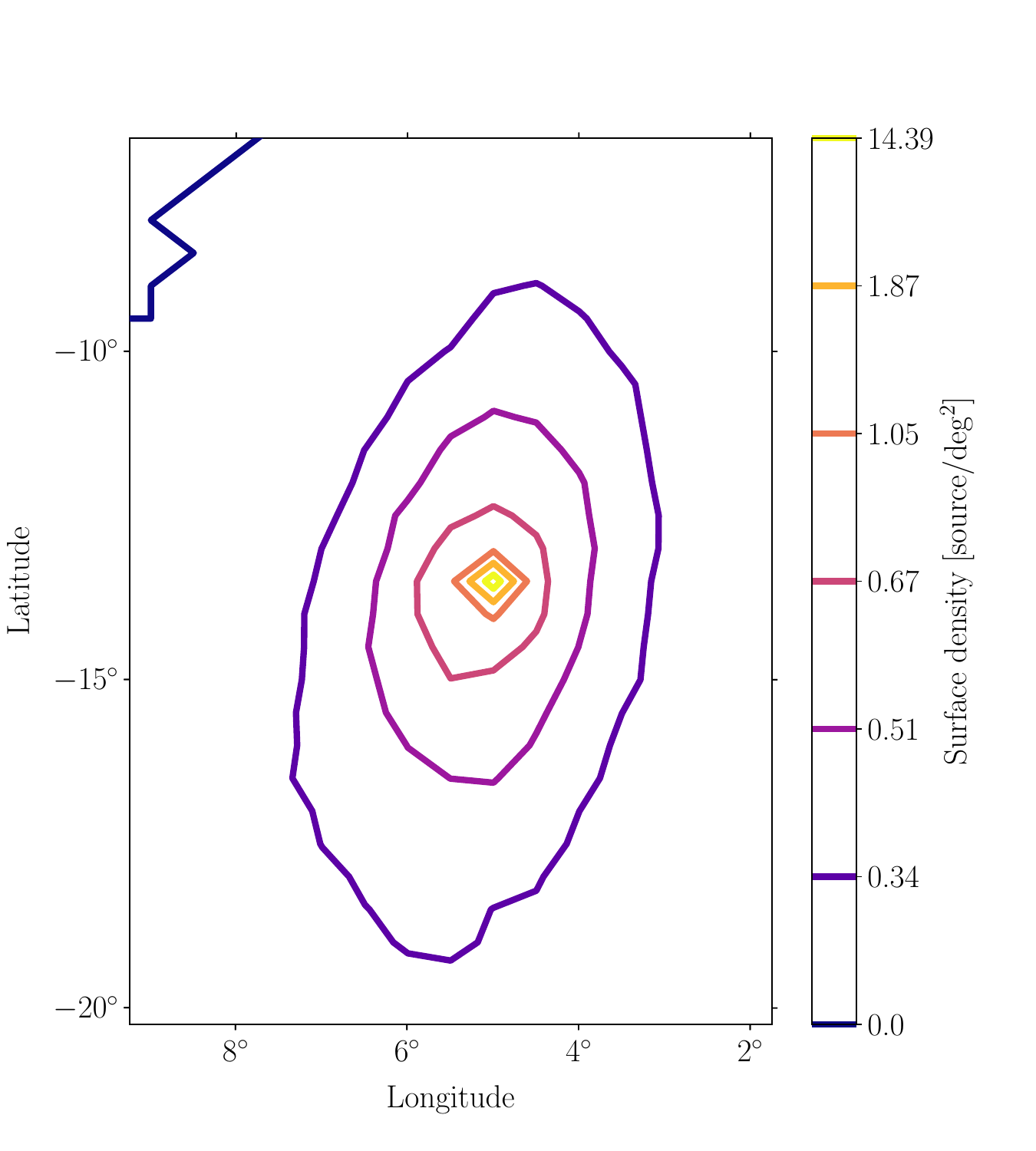}
    \caption[]{Surface density of radio-bright MSPs in Sgr dSph per deg$^2$. Based on this map, is more likely to observe MSPs towards the core of the galaxy, approximately within $\sim 5$ deg$^2$.}
    \label{fig:SurfDens}
\end{figure}
To determine the energy intensity in gamma-rays of the Sgr dSph, we have used the best-fit spectrum as reported in \citet{Crocker2022}. This spectrum corresponds to a power law with an exponential cutoff of $\gamma_\mathrm{MSP}$ = 1.117 and $E_\mathrm{cut,e^{\pm}}$ = 0.5 TeV.

We calculated the total energy intensity by integrating the energy spectrum ($E dN/dE$) of the Sagittarius dwarf spheroidal galaxy, starting from 300 MeV. For creating the surface density map, we used the distribution of RR Lyrae in Sagittarius dwarf spheroidal galaxy from Gaia Data Release 2 \footnote{\url{https://gea.esac.esa.int/archive/}}. Since MSPs and RR Lyrae are old stellar populations, they likely share a similar distribution and dynamic history. However, the stellar distribution data has a noticeable step between stellar bins. To ameliorate this, we interpolated across neighboring stellar bins (see Appendix \ref{appendix:distribution}). We calculated the solid angle within the contours that contain a specific percentage of RR Lyrae stars. Then, we multiplied the energy intensity by the RR Lyrae fraction within its given solid angle.

Finally, we constructed a map by dividing the energy intensity by the ratio $\mathcal{R}_\mathrm{rb}^\gamma$, and then multiplying it by $4\pi d_\mathrm{Sgr}$, where $d_\mathrm{Sgr}$ represents the distance to Sgr dSph, which is 26.5 kpc. The outcomes are shown in Fig.~\ref{fig:SurfDens}. For a solid angle of $\Omega_\mathrm{Sgr} \simeq 9.6\times 10^{-3}$ sr, which includes most of the galaxy, we found a surface density of $0.57 \text{deg}^{-2}$.

\section{Radio Telescope Sensitivity}\label{sec:RadioTelescopes}

The radiometer equation can be used to determine a radio telescope's sensitivity to detect pulsations originating from pulsars. This equation enables calculating the minimum flux density required to detect a pulsar with a $10 \sigma$ significance level. Essentially, the radiometer equation (as described in \citealt{Dewey_1985}) provides a value that acts as a threshold --- if the pulsar's flux density exceeds this value, it will be detectable.

The radiometer equation can be expressed as follows:

\begin{equation}
    \label{eq:radiometer}
    S_{\nu,\text{rms}} = \frac{10 (T_\text{sky}+T_\text{rx})}{G\sqrt{t_\text{obs} \Delta\nu n_\text{p}}}\left(\frac{W_\text{obs}}{P -W_\text{obs}}\right)^{1/2},
\end{equation}
where $T_\text{sky}$ and $T_\text{rx}$ are the sky and receiver temperatures (in K), $G$ is the telescope gain (in K/Jy), $n_\text{p}$ is the number of polarizations, $\Delta\nu$ is the frequency bandwidth (in MHz), $t_\text{obs}$ is the integration time (in s), $W_\text{obs}$ is the observed pulse width and $P$ is the pulsar spin period. Below, we describe how we estimate the sensitivity ($S_{\nu,\text{rms}}$) of our radio pulsation searches.

\subsection{Telescope Information}\label{subsec:Telescope_info}

In this section, we will discuss the various parameters relevant to pulsar detection, including the receiver temperatures $T_\text{rx}$, the telescope gain $G$, and the frequency bandwidth $\Delta\nu$. Technical information about MeerKAT used in this study can be found in the External Service Desk Knowledge Base\footnote{\url{https://skaafrica.atlassian.net/wiki/spaces/ESDKB/pages/277315585/MeerKAT+specifications}}. It is worth noting that the Square Kilometer Array (SKA) is not yet ready to perform observations. However, information about the system baseline can be found on the Square Kilometer Array Observatory webpage\footnote{\url{https://www.skao.int/en/science-users/118/ska-telescope-specifications}}.

To calculate the gain sensitivity, we need to account for the decrease of the gain by a factor of two towards the Full Width at Half Maximum (FWHM) edge of the telescope beam. We calculate the antenna gain using the system-equivalent flux density (SEFD), which is the flux we would observe of a source with a temperature equivalent to that of the receiver. For MeerKAT and SKA, we assume a receiver temperature of 25K. We will consider 4096 channels to divide the incoming frequency band affected by dispersion in the interstellar medium, using the information available for the L-band (856–1712 MHz). The L-band has a total bandwidth of 856 MHz, but this does not take into account any potential radio frequency interference during observations. For SKA we will use SKA1-Mid band 2 (950-1760 MHz), which corresponds to a bandwidth of 810 MHz. The sampling time $\tau_\mathrm{samp}$ for gathering data is chosen based on previous MSP searches using MeerKAT (e.g., \citet{Zhang_2022, Ridolfi2021}), and we set $\tau_\mathrm{samp} = 76 \mu s$ for 4096 channels. Since the SKA array is not yet available, we will use some of the same parameters as MeerKAT for SKA specifications.

\begin{figure}
    \centering
    \includegraphics[scale=0.46]{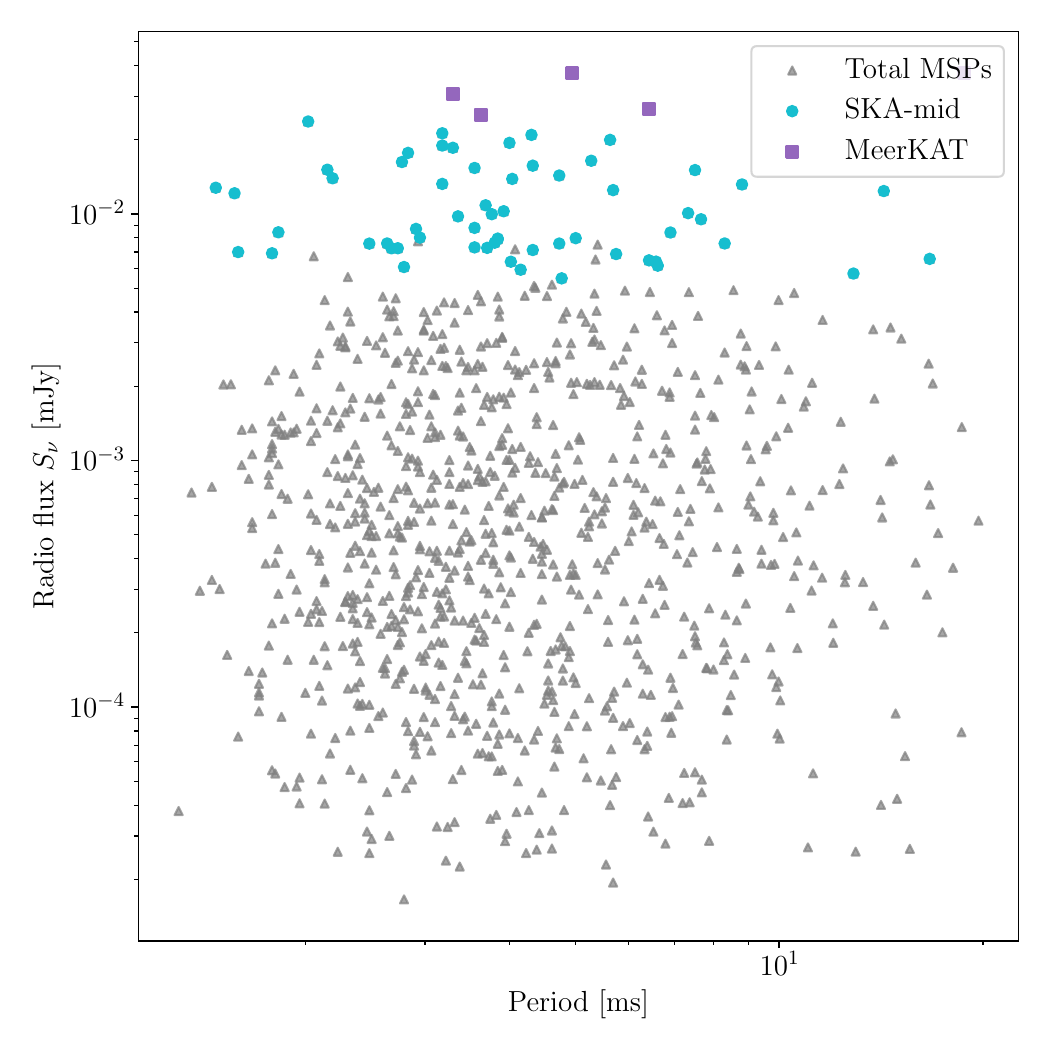}
    \caption[]{Radio flux density at 1.4GHz versus period of simulated Sgr dSph MSPs. Using one Monte Carlo realization, we can picture the amount of MSPs that can be detected by MeerKAT and SKA. The grey triangles are the total simulated pulsars, the cyan circles represent the pulsars that can be detected using SKA (63 in this case), and the purple squares show the MSPs that can be detected by SKA and MeerKAT.}
    \label{fig:flux_vs_period}
\end{figure}

\subsection{Source Information}\label{subsec:SourceInfo}

We must sample specific parameters dependent on our location in the Galaxy. Firstly, we estimate the temperature of the sky, denoted as $T_\mathrm{sky}$, by using the most recent model for the diffuse Galactic radio emission from \citet{Zheng_2017}, which can be used to obtain sky maps ranging from 100 MHz to 100 GHz. 
Secondly, to calculate the pulse width that we observe from the pulsar, we need to understand an important phenomenon that takes place between the source and our observation point: as radio waves from a pulsar travel through an ionized interstellar medium, they experience a delay in their arrival times. This delay in arrival times is inversely proportional to the observing frequency, and the constant of proportionality is known as the dispersion measure (DM) \citep{LorimerKramer2004}. This delay causes the received signal to be broad, which reduces the signal-to-noise (S/N) ratio. We note that to detect pulsations, the observed pulse width, denoted as $W_\text{obs}$ (in ms), should be smaller than the source period $P$ (in ms). Lastly, the observed pulse width $W_\text{obs}$ is determined by the intrinsic pulse width and the effects that smear the observed pulse profile:

\begin{equation}
    \label{eq:W_obs}
    W_\text{obs} =\left( (w_\text{int}P)^2 + \tau_\text{DM}^2  +\tau_\text{scatt} +\tau_\text{samp}^2 + \tau_{\Delta\text{DM}}^2 \right)  ^{1/2},
\end{equation}
where $w_\text{int}$ is the pulsar's intrinsic pulse width (0.1 for MSPs, see \citet{LorimerKramer2004}), $\tau_\text{DM}$ is the dispersive smearing across individual frequency channels, $\tau_\text{scatt}$ is the temporal smearing due to interstellar scattering, $\tau_\text{samp}$ represents the time for sampling data, and $\tau_{\Delta \text{DM}}$ is the smearing due to the deviation of a pulsar's true DM from the nominal DM. 
Since the DM step is small, we will neglect $\tau_{\Delta DM}$.

\begin{figure}
    \centering
    \includegraphics[scale=0.5]{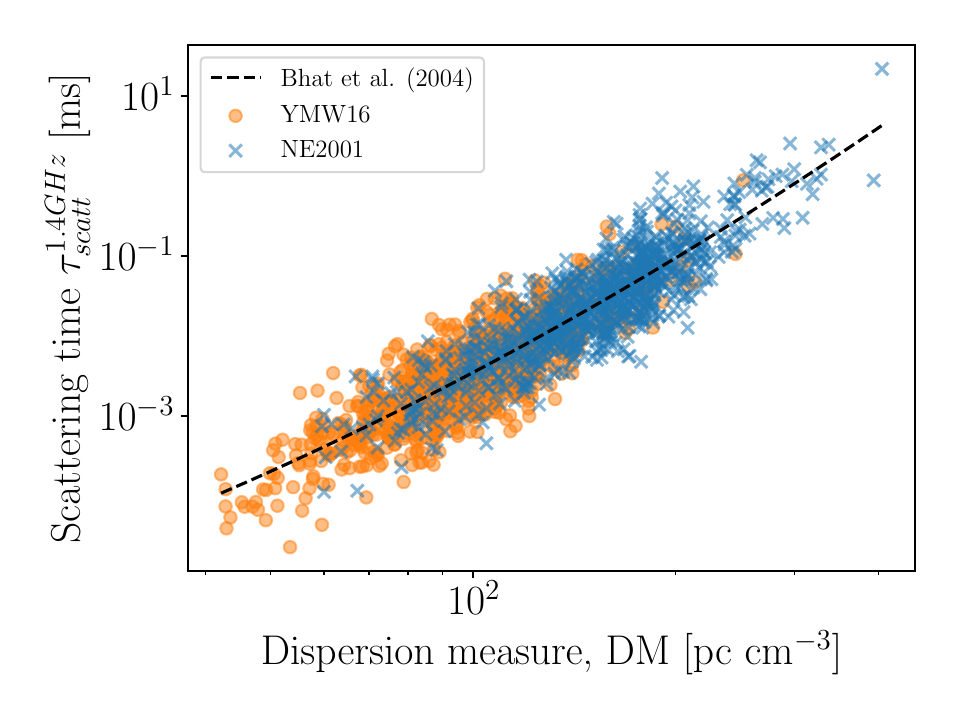}
    \caption[]{Sampled values for scattering time versus their dispersion measure. Blue dots are the values sampled using the DM model NE2001, and the orange dots are sampled based on the YMW16 model. The black dashed line is the empirical formula from \citet{Bhat_2004}.}
    \label{fig:scatt_vs_DM}
\end{figure}

Typically the intra-channel smearing $\tau_\text{DM}$ is the smearing due to finite DM step size in the search. 
Assuming that the channel bandwidth $\Delta\nu_{chan}$ is much smaller than the reference frequency $\nu$, $\tau_\text{DM}$ is given by \citet{Hessels_2007} as follows:
\begin{equation}
    \tau_\text{DM} = 8.3 
    \left(\frac{\text{DM}}{\text{pc cm}^{-3}}\right)
    \left(\frac{\Delta\nu_{\text{chan}}}{\text{MHz}}\right)
    \left(\frac{\nu}{GHz}\right)^{-3}.
\end{equation}
We do not have prior knowledge of the dispersion measure of the putative MSPs in the Sgr dSph. However, we can sample from a reasonably likely distribution of these MSPs. As mentioned before, we determine the position of MSPs in Sgr dSph based on the RR Lyrae distribution. In particular, we sample the distance from a normal distribution with a mean of 26.5 kpc and a standard deviation of 1 kpc. Using the model for free-$e^{\pm}$ by \citet{Cordes_&_Lazio2002}(NE2001) and \citet{Yao_2017} (YMW16), we can calculate the DM for any line-of-sight and distance of the source. In Figure \ref{fig:scatt_vs_DM}, we can see each model's sampled dispersion measure values. The values obtained from NE2001 are relatively higher than the others. Therefore, the results from this model can provide us with a more conservative constraint. None of the models can estimate the DM in the intergalactic medium or in Sgr dSph. However, considering only the DM from the Milky Way in the Sgr dSph is less likely to significantly affect the results due to the absence of gas and young stellar systems there.

The broadening of pulsar signals is caused by fluctuations in the electron density and distribution of scattering material along the line of sight. This effect can be measured by a timescale ($\tau_\text{scatt}$) and depends on factors such as the dispersion measure, line of sight direction, and frequency. An empirical formula proposed by \citet{Bhat_2004} helps quantify this effect:
\begin{equation}
    \log{\tau_\text{scatt}} = a + b(\log{\text{DM}})+c(\log \text{DM})^2 - \alpha \log \nu,
\end{equation}
where $\nu = 1.4$GHz, $a=-6.4$, $b=0.154$, $c=1.07$ and $\alpha=4.4$.
Here, we use a lognormal distribution with a mean $\mu=\log_{10}{\tau_\text{scatt}}$ and a variance $\sigma=0.8$ to sample the values for $\tau_\text{scatt}$.

In a recent study,  \citet{Liu_2023} analyzed the distribution of spin periods of millisecond pulsars (MSPs) with periods ranging from 0.6 ms to 20 ms. The study focused on three different samples of MSPs: sample A, which included 260 MSPs observed at 1400 MHz in the galactic field; sample B, which included all 473 radio MSPs in the galactic field; and sample C, which included 273 MSPs found in globular clusters. We will focus on the results of sample A for this study since we are interested in finding MSPs in the Sgr dSph region. The best fit for sample A was a cutoff model represented by $\mathcal{N} = H(P-P_\mathrm{min}) \log\mathcal{N}(\mu',\sigma')$, where $H(P-P_\mathrm{min})$ is a Heavyside function, and $\log\mathcal{N}(\mu',\sigma')$ is a log-normal distribution. The values of $P_\mathrm{min}$, $\mu$, and $\sigma$ were determined to be 1.43 ms, 1.44, and 0.6, respectively.

\section{Radio Sensitivity to MSPS in the Sagittarius Dwarf} \label{sec:Results}

\begin{figure*}
    \centering
    \includegraphics[scale=0.7]{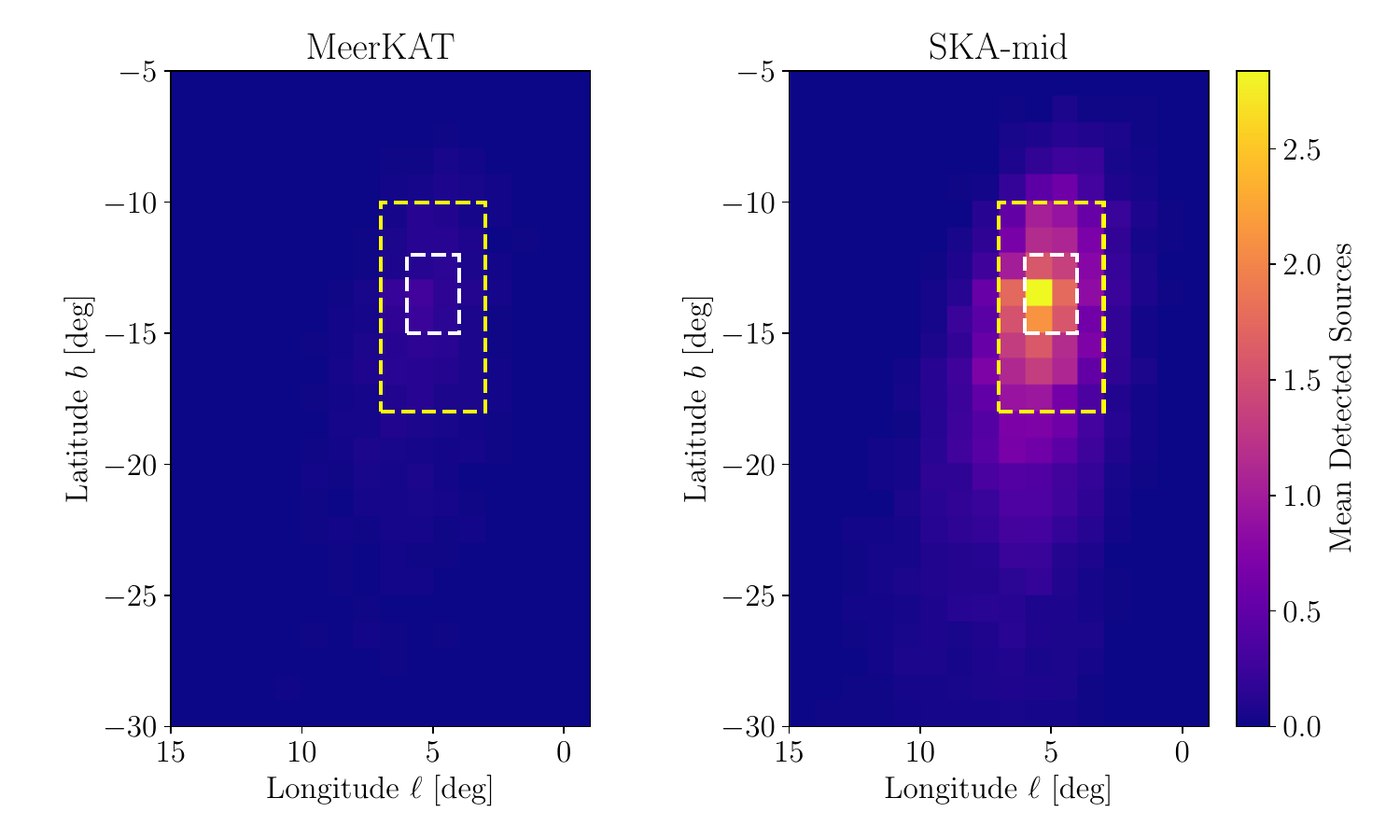}
    \caption[]{Sky map of the mean of detected sources by MeerKAT (right panel) and SKA (left panel), using the piece-wise period distribution proposed in this work and model NE2001. For each trial was calculated the sources detected by using an integration observing time of 20 min per pointing ($1^\circ \times 1^\circ$). 
    The dashed white line shows the region that would take 2 hours in total to observe and the yellow dashed line shows the region that would take $\sim 11$ hours.
    SKA-mid is the most promising telescope to observe this region. In both cases, we see a bigger density of sources in the galaxy's core.}
    \label{fig:results_ska_mrkt}
\end{figure*}
In this section, we explore the potential for present and upcoming radio telescopes to detect MSPs in the Sgr dSph galaxy. We also estimate the minimum number of MSP detections required to provide irrefutable evidence of the existence of an MSP population in the Sgr dSph galaxy.

We performed Monte Carlo simulations, carrying out 300 trials in which we sampled values for the parameters indicated in equation \ref{eq:radiometer}. In each trial, we sampled the information of $986$ MSPs, i.e. the number of radio MSPs obtained in Sec. \ref{sec:method_GC_to_Sgr}. For each MSP, we sample the position distance and all properties referred in Sec. \ref{sec:RadioTelescopes}

Since there are patches in our simulations where we do not detect any MSPs, some sectors of the Sgr dSph might not be well described by a normal distribution. Thus, we use quantiles to reflect the actual spread of the number of sources, regardless of the specific shape of their spatial distribution. The results presented here correspond to the 50th quantile, with the lower error being the 10th and the higher being the 90th quantile.

During our study, we utilized an integrated observation time of 20 minutes per pointing with a size of $1^\circ \times 1^\circ$. After conducting 300 trials, we obtained that MeerKAT would be able to observe $5^{+3}_{-2}$ MSPs in the Sgr dSph, while SKA would be able to observe $54^{+10}_{-8}$ MSPs, for the observation times nominated in Table~\ref{tab:results_per_telescope}. It is worth noting that the observation time $t_\mathrm{obs}$ may vary depending on the source's declination \citep{Hessels_2007}. 

We have also evaluated the feasibility of observing the entire region of the Sgr dSph using radio telescopes. However, due to the challenging nature of this task, which would require approximately 70 hours with MeerKAT, we have decided to focus on the areas that are more likely to have pulsars within the dwarf. The core of the Sgr dSph, located within a longitude of $3^\circ$ to $7^\circ$ and a latitude of $-10^\circ$ to $-18^\circ$, is the most promising region for analysis~\footnote{Note that this angular scale also encompasses the globular cluster M54 but it extends to a significantly wider angular scale beyond the globular cluster.}. According to MeerKAT's specifications, we would need at least two hours of observation time to detect one MSP. In the same observation time, the SKA-mid has the potential to detect $11\pm3.2$ MSPs. Interestingly, by observing the Sgr dSph core, which would require 10.6 hours, we can detect $3\pm1.8$ MSPs using MeerKAT and $36\pm6.0$ with SKA.

\begin{table}
    \centering
    \begin{tabular}{|l c c c|}
    \hline
    Sky Surface & Region & NE2001 & YMW16\\
    \hline
    \ & $2^\circ \times 3^\circ$ & $1^{+1}_{-1 }$ & $1^{+1}_{-1}$ \\
    MeerKAT & $4^\circ \times 8^\circ$ & $3^{+2}_{-2 }$ & $3^{+3}_{-2}$ \\
    \ & Total & $5^{+3}_{-2}$ & $6^{+3}_{-3}$\\
    \hline
    \ & $2^\circ \times 3^\circ$ & $9^{+5}_{-3 }$ & $9^{+5}_{-3}$ \\
    SKA-mid & $4^\circ \times 8^\circ$ & $30^{+8}_{-6}$ & $32^{+7}_{-7}$\\
     \ & Total & $54^{+10}_{-8}$ & $57^{+9}_{-9}$\\
    \hline
    \end{tabular}
    \caption[]{Number of MSPs detectable using MeerKAT and SKA with an integration time of 20 mins per pointing (where one pointing is $1^\circ \times 1^\circ$). The results shown here correspond to the quantile 50 and the lower error is quantile 10 and the higher 90.}
    \label{tab:results_per_telescope}
\end{table}

The above results are based on Model 3 in \citet{Bagchi_2011} study. However, when simulations are conducted assuming Model 1 in that study, no MSPs can be detected with MeerKAT. On the other hand, if we observe the entire region, SKA can detect a total of 6 MSPs with an uncertainty range of $\pm 3$. For Model 2, MeerKAT can detect 4 MSPs with an uncertainty range of $\pm 2$, while SKA can detect 46 MSPs with an uncertainty range of $\pm 9$. These results are based on the dispersion model NE2001, and the number of radio-bright MSPs considered is 986.

\subsection{Foreground pulsars}\label{subsec:foreground_pulsars}

\begin{figure*}
    \centering
    \includegraphics[scale=0.45]{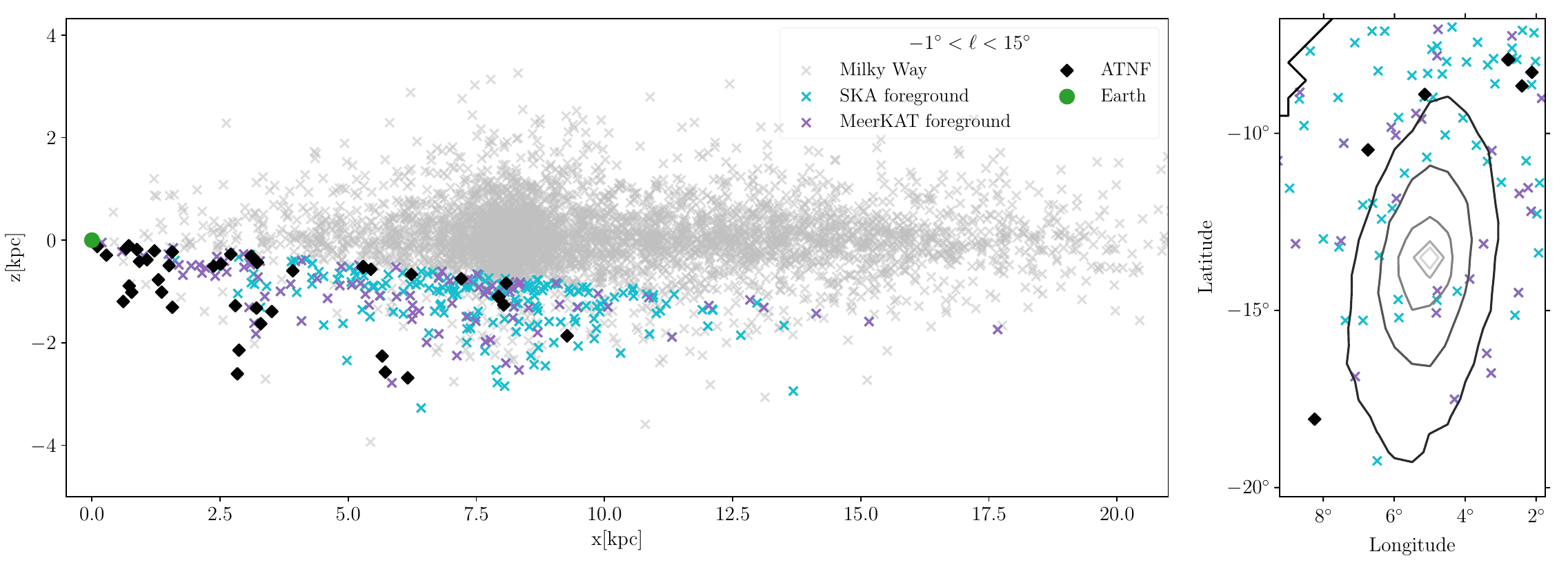}
    \caption[]{\textit{Left:} Map of pulsars in the Milky Way within  $ -1^\circ<\ell<15^\circ $. The grey crosses correspond to one of the stellar distributions simulated for the MW, the Earth is represented as a green circle. The rest of the pulsars shown on the plot are on the line of sight to Sgr dSph. 
    The purple crosses correspond to the MSPs that could be detected with MeerKAT and SKA, while the cyan crosses represent the MSPs that could only be detectable using SKA. The black diamonds are known pulsars based on the ATNF Catalog\footnote{Data published by 25th January 2024.}. \textit{Right:} Contour of the stellar distribution of Sgr dSph. The skymap shows pulsars that we could potentially detect that are found in the line of sight of the core of Sgr dSph.}
    \label{fig:foreground_pulsars}
\end{figure*}

The direction towards the Sgr dSph galaxy is near the Galactic bulge and passes through the Galactic disk. This poses a risk of mistakenly attributing an MSP that belongs to either the disk or the bulge to the Sgr dSph. However, as the MSPs we could potentially discover in the Sgr dSph are located far enough away, we could differentiate between the MSPs belonging to the disk or bulge and those belonging to the Sgr dSph by estimating their dispersion measure distance.

We simulate MSPs in the galactic bulge and disk to estimate the number of detectable foreground sources. In the bulge, we use an inverse power-law function of the galactocentric distance $r$, with an index of $\Gamma=2.56$  and $r_\mathrm{cut}>1.1$kpc, as reported in \citet{Calore_2016}. We also adopt a cut-off at 3 kpc, as suggested by \citet{Calore_2016}. For the galactic disk, we assume that the MSP population follows an exponential disk distribution. Specifically, we simulate the radius $r$ and the height $z$ as an exponential distribution where the scale radius is 5 kpc, and the height scale is 0.5 kpc \citep{Faucher_Gigu_re_2010}. To estimate the sensitivity, we use Model 3 from \citet{Bagchi_2011} to sample the radio flux, following the same procedure we used in the Sgr dSph MSPs.

The results of our simulations are presented in Figure \ref{fig:foreground_pulsars}. Detectable foreground MSPs using different instruments and all pulsars detected in the line of sight based on the ATNF catalog are also shown. Our findings suggest that if we observe a small region around the core of the dwarf (measuring $2^{\circ} \times 3^{\circ}$), we could potentially detect 1 and 4 MSPs on the Milky Way using MeerKAT and SKA, respectively. However, if we expand our observation to include the entire core ($4^{\circ} \times 8^{\circ}$), MeerKAT could detect up to 10 MSPs, and SKA could detect up to 25 MSPs. These values remain relatively consistent across different simulation models. It should be noted that we did not run a Monte Carlo simulation in this case, so the results of possible MSPs detectable in the MW galaxy might vary. Nevertheless, this gives us an idea of sources that might contribute to the gamma-ray excess in the cocoon.

\subsection{Minimum Number of MSPs necessary to confirm the MSPs theory in the Sgr dSph}

In Table~\ref{tab:results_per_telescope}, we have calculated the average number of MSPs detected using upcoming telescopes in the Sgr dSph through Monte Carlo sampling. In this section, we aim to determine the minimum number of MSPs that must be observed in this direction to confirm that the gamma-ray emission from this region is indeed attributed to a new MSP population in the Sgr dSph.

We used Monte Carlo simulations to compare the detectable pulsars in the Milky Way and Sgr dSph. The simulations were conducted similarly in both cases, with the only difference being the use of different distance parameters for each object. This led to changes in the dispersion measure and scattering time. Our findings suggest that with MeerKAT, we may be able to detect $1\pm1$ MSPs in the Milky Way toward the line of sight of the dwarf, and with SKA, we could potentially detect $3\pm1$ MSPs. This result is within the $2^\circ \times 3^\circ$ region that we see within the white dashed rectangle in Figure~\ref{fig:results_ska_mrkt}.

To ensure that the gamma-ray emission originates from an MSP population in the Sgr dSph with a 99.7\% confidence level (or $3\sigma$ statistical significance), we conduct a likelihood-ratio-test evaluated with the Asimov-data-set \citep{Cowan2011, Calore_2016}. The Asimov data set serves as a model of the data we would expect to observe if the alternative hypothesis were true. In this case, the ``Asimov data set'' is denoted by $c_i^{\mathrm{Alt}} = \zeta (\mu_i^{\mathrm{Sgr}} + \mu_i^\mathrm{MW})$, where
$\mu_i^{\mathrm{Sgr}}$ and $\mu_i^{\mathrm{MW}}$ are the pulsars detected in different $1^\circ \times 1^\circ$ sky patches in the Sgr dSph and Milky Way, respectively. We solve for the scale factor, $\zeta$, by requiring that $-2\ln(\mathcal{L_\mathrm{null}}/\mathcal{L_\mathrm{alt}}) = 9$. For the null hypothesis, we only consider detections in the Milky Way, and we assume that the likelihood function  $\mathcal{L}_\mathrm{null}$ is given by a normal distribution with $\mu = \mu_i^{\mathrm{MW}}$ and $\sigma = \sigma_i^{\mathrm{MW}}$.
For the alternative hypothesis we consider the detections on the Milky Way and Sgr dSph, $\mathcal{L}_\mathrm{alt}$ is also described by a normal distribution with $\mu = (\mu_i^{\mathrm{Sgr}} + \mu_i^\mathrm{MW})$ and $\sigma = (\sigma_{i,\mathrm{MW}}^2 + \sigma_{i,\mathrm{Sgr}}^2 )^{0.5}$.

When observing with SKA, we obtained a scale factor $\zeta = 0.7$. This means we can identify an MSP population at the 99.7\% confidence level by detecting at least 6 MSPs in the Sgr dSph and 2 in the Milky Way in a two-hour observation.

We note that the population of MSPs in the Sgr dSph can be differentiated from foreground MSPs in the Milky Way based on their dispersion measure. Figure~\ref{fig:foreground_pulsars} illustrates that the dispersion measure of MSPs in the Milky Way has a notably lower value than those in the Sgr dSph.

\section{Summary and Conclusions}\label{sec:DiscussionConclussions}

The nature of the Fermi Cocoon remains a mystery. However, an exciting possibility has emerged: an undiscovered population of millisecond pulsars in the Sgr dSph~\citep{Crocker2022}. The discovery of such MSPs using multi-wavelength observations could decisively solve this astrophysical puzzle.

In this article, we evaluated the sensitivity of MeerKAT and SKA radio telescopes to a putative population of MSPs in the Sgr dSph. To estimate the number of radio MSPs in the dwarf, we used a phenomenological model using a methodology similar to one implemented by \citet{Calore_2016}. This allowed us to calculate the ratio of gamma-ray luminosity to the number of radio sources in globular clusters, which we then extrapolated to the Sgr dSph. We assumed that both the globular clusters and the dwarf have similar dynamical histories, and we rescaled the stacked globular clusters to the dwarf's distance. 
In our future work, we will conduct population synthesis simulations similar to the one described in \citet{Gautam_2022}. These simulations will enable us to explore the effect of this assumption on our sensitivity estimates.

Based on the MSPs luminosity model proposed by \citet{Bagchi_2011}, we found that the number of radio MSPs that could account for the gamma-ray emission in the Sgr dSph is estimated to be  $986 \pm 555$, out of which $20 \pm 12$  are radio-bright MSPs (with a flux density $S_\mathrm{1.4 GHz} >0.01$ mJy). It is worth noting that in \citet{Crocker2022}, the number of MSPs in the Sgr dSph was calculated using an Accretion Induced Collapse model, which resulted in a value of approximately 650 MSPs. Interestingly, this value is within the $1\sigma$ error of the total number of radio MSPs we obtained for Model 2 and Model 3 (see Table~\ref{tab:results_per_model}).

We demonstrated that if we observe the densest region of Sgr dSph ($2^\circ \times 3^\circ$) divided in 6 pointings of $1^\circ \times 1^\circ$, we need a total time of 2 hours to observe 1 MSP using MeerKAT. In contrast, we obtained that SKA could detect $9^{+5}_{-3}$ under the same conditions. However, it is important to note that our SKA forecast is based on our current understanding of the instrument; we foresee that this will need to be updated once the telescope is better understood.

Our sensitivity estimates remain broadly consistent whether we use the dispersion model NE2001 or YMW16 (as noted in Table~\ref{tab:results_per_telescope}). In the particular case of the core region of the galaxy, we observed that the DM values are higher when we use the NE2001 model, making it a more conservative estimate. It is important to note that our analysis is limited to electron scattering within the Milky Way, and we do not have any measurements for scattering that occurs beyond it (at least not towards the Sgr dSph). If we observe foreground pulsars, we can use the dispersion measure towards them to estimate their distance and determine if they belong to the Milky Way or Sgr dSph. 

MSPs emit X-rays through thermal emission from heating in the polar caps and non-thermal emission resulting from the particle acceleration in the pulsar magnetosphere. There can also be emission from interactions between the pulsar wind and the material from a binary companion, if present \citep{Bogdanov_2017}. In our upcoming research, we will explore the possibility of discovering a potential new population of MSPs in the Sgr dSph using X-ray observatories.

\section*{Acknowledgements}

We thank Shin'ichiro Ando, Francesca Calore, and Roland Crocker for fruitful discussions. OM was supported by the U.S. National Science Foundation under Grant No 2418730. In this work, we extensively used publicly available software Astropy \citep{astropy_2018}, PyGEDM \citep{Pygedm_2021}. General python packages: NumPy \citep{harris2020array}, Matplotlib \citep{Hunter:2007}, HEALPix \citep{Gorski2005, Zonca2019}. 
\section*{Data Availability}
 
The data required to reproduce the results in this article are available under reasonable request to the authors.



\bibliographystyle{mnras}
\bibliography{references} 



\appendix

\section{Stellar Distribution}\label{appendix:distribution}
For the stellar distribution of Sgr dSph we assumed that MSPs will follow the same distribution of RR Lyrae. This is a reliable assumption since both objects are old hence is highly likely they share the same distribution nowadays.
To sample the stellar distribution of Sgr dSph we use the map of RR Lyrae detected in this region. However, as we can see in the data has a noticeable step between each bin. To create a realistic distribution we interpolate the data of the 2D bins using a bivariate spline approximation over a rectangular mesh grid included on the \verb|SciPy| package.


\bsp	
\label{lastpage}
\end{document}